\documentclass{aa}  

\usepackage{subcaption}
\usepackage{multirow}
\usepackage{graphicx}
\usepackage{txfonts}
\begin{document} 

   \title{The rise and fall of little red dots could be driven by the environment}

   \titlerunning{The rise and fall of little red dots could be driven by the environment}

   \authorrunning{Rosa M. M\'erida et al.}
    \author{Rosa M. Mérida\inst{1} \and Gaia Gaspar\inst{1,2} \and Yoshihisa Asada \inst{1,3,4,5} \and Marcin Sawicki\inst{1} \and Kiyoaki Christopher Omori\inst{1} \and\\ Chris J. Willott\inst{6} \and Nicholas S. Martis \inst{7} \and Adam Muzzin \inst{8} \and Ga\"el Noirot \inst{9} \and Gregor Rihtar\v{s}i\v{c} \inst{7} \and\\ Ghassan T. E. Sarrouh \inst{8} \and Roberta Tripodi \inst{10}}

   \institute{
                Institute for Computational Astrophysics and Department of Astronomy and Physics, Saint Mary's University, 923 Robie Street, Halifax, NS B3H 3C3, Canada\\
              \email{Rosa.MeridaGonzalez@smu.ca}
         \and
             Observatorio Astronómico de Córdoba, Universidad Nacional de Córdoba, Laprida 854, X5000, Córdoba, Argentina
        \and
             Dunlap Institute for Astronomy and Astrophysics, 50 St. George Street, Toronto, Ontario, M5S 3H4, Canada
        \and
             Waseda Research Institute for Science and Engineering, Faculty of Science and Engineering, Waseda University, 3-4-1 Okubo, Shinjuku, Tokyo 169-8555, Japan
        \and
             Department of Astronomy, Kyoto University, Sakyo-ku, Kyoto 606-8502, Japan
             \and
        National Research Council of Canada, Herzberg Astronomy \& Astrophysics Research Centre, 5071 West Saanich Road, Victoria, BC, V9E 2E7, Canada
        \and
        Faculty of Mathematics and Physics, Jadranska ulica 19, SI-1000 Ljubljana, Slovenia
        \and
        Department of Physics and Astronomy, York University, 4700 Keele St. Toronto, Ontario, M3J 1P3, Canada
        \and
        Space Telescope Science Institute, 3700 San Martin Drive, Baltimore, Maryland 21218, USA
        \and
        INAF - Osservatorio Astronomico di Roma, Via Frascati 33, Monte Porzio Catone, 00078, Italy
        }

   \date{Received September 15, 1996; accepted March 16, 1997}

  \abstract
   {The little red dot (LRD) paradigm comprises three main unknowns that are intrinsically connected: (1) What is the nature of these sources? (2) How do they form? (3) How do they evolve? Larger spectroscopic samples and high-resolution data are needed to delve deeper into the mechanisms governing these sources. Understanding their formation and evolution requires identifying the rise and fall of the key features that characterize these systems, such as their compactness and ``V''-shaped spectral energy distributions. In this work, we present a galaxy system nicknamed ``The Stingray" that was identified in the Canadian NIRISS Unbiased Cluster Survey (CANUCS). This group contains three sources at $z_{\mathrm{spec}} = 5.12$, including an active galactic nucleus (AGN), a Balmer break galaxy, and a star-forming satellite. The latter resembles a building blocks system in which interactions boost stellar mass and black hole mass growth beyond what is expected from secular processes alone. The AGN in this system exhibits features indicative of a transitional object, bridging a normal AGN and an LRD phase. These features include a blue rest-frame UV slope, compact size, and a broad H$\alpha$ line (all of which are characteristic of LRDs), but a flatter rest-frame optical slope compared to that observed in LRDs. The features in this source point to the emergence or fading of an LRD, potentially triggered by environmental effects.}


   \maketitle
%

\section{Introduction}
\label{sec:intro}

Two years after the discovery of the so-called little red dots (LRDs; \citealt{Barro2024}, \citealt{Greene2024}, \citealt{Labbe2023}, \citealt{Matthee2023}) by the JWST \citep{Gardner2023}, our understanding of their formation, evolution, and properties remains highly incomplete.
Consistent features across all studies of these intriguing sources are their compact nature, the presence of broad Balmer emission lines, and a distinctive ``V''-shaped spectral energy distribution (SED). This SED is characterized by a nearly flat-to-blue rest-frame UV continuum, followed by a steep optical slope. However, identifying the complete sample of such sources remains challenging: photometric selections can be contaminated by emission lines that mimic steep spectral slopes \citep{Hainline2024arxiv}. Moreover, they can also overlook objects that reveal ``V''-shaped continua only through spectroscopy \citep{Hviding2025}.    

Different combinations of models have been proposed to explain the nature of these objects, including active galactic nuclei (AGNs), stellar emission, and composite models (e.g., \citealt{Greene2024}, \citealt{Perez-Gonzalez2024}, \citealt{Zhengrong2024}, \citealt{Merida2025}, \citealt{Tripodi2024}). None of these models has fully reproduced the bulk of the LRD population and their challenging properties, which include a lack of X-ray detections \citep{Ananna2024}, an absence of emission from a dusty torus at longer wavelengths \citep{Setton2025}, and a lack of variability \citep{Zhang2024}.
As a result, more exotic solutions have been proposed, such as the black hole star (BH*) model (e.g., \citealt{deGraaf2025}, \citealt{Inayoshi2024}, \citealt{Ji2025}, \citealt{Naidu2025}, \citealt{Rusakov2025}) or binary massive black holes \citep{Inayoshi2025}. However, larger spectroscopic samples and high-resolution observations are still required to further investigate these hypotheses.

Another key aspect of the LRD paradigm is the emergence and evolution of these sources. Although candidates for $z\sim0$ LRD analogs now exist (see \citealt{RLin2025,Lin2025}, \citealt{Ji2025b}), the population of these sources exhibits a dramatic drop in number density at $z\sim 4$ ($1-2$ dex from $z\sim 4$ to $z \sim 2$; \citealt{Kocevski2025}, \citealt{Inayoshi2025b}, \citealt{Ma2025}, \citealt{Zhuang2025}). The evolutionary pathway followed by these sources has not yet been fully elucidated. 

According to \citet{Billand2025}, LRDs represent a phase in galaxy evolution, which implies that transition stages must exist in which galaxies exhibit LRD features while simultaneously developing new characteristics. These authors also note that cold accretion may increase stellar mass ($M_\star$), causing LRDs to lose their genuine ``V''-shape and compactness as these stars form and populate the galaxy outskirts. 
In that same work, mergers are also proposed as channels for the disruption of the LRD phase. 

Observations already reveal LRDs in dense environments, including LRDs in pairs (\citealt{Tanaka2024}, \citealt{Merida2025}), LRDs in galaxy overdensities (\citealt{Labbe2024environment}, \citealt{Rinaldi2024}, \citealt{Schindler2025}), and star clusters or nebular gas near LRDs \citep{Chen2025}. 
In \citet{Merida2025} we present evidence of a synchronous evolution of an LRD pair in $z\sim7$ and highlight the key role that the environment plays in triggering and possibly quenching these objects. 

Different channels likely exist for the formation and destruction of LRDs, and environmental effects should be considered as potential drivers of some of these channels.
To understand how LRDs are triggered and disappear, it is essential to investigate the properties of transitional objects. 
The presence of these sources can be linked both to the end of an LRD phase and to the emergence of one or more LRD phases during the lifetime of a galaxy.

In this work, we report the discovery of a possible transitional object bridging normal AGNs and LRDs, embedded in a galaxy system nicknamed The Stingray at $z_{\mathrm{spec}}=5.12$. This group contains a potential transitional LRD (tLRD), a Balmer break galaxy (BBG), and a star-forming satellite. All three sources show signs of coordinated star formation histories (SFHs), resembling the building blocks system (BBS; Asada et al. in prep). In such systems, stellar mass ($M_\star$) growth can be dramatically enhanced by interaction-triggered bursts of star formation.

Throughout this work, we assume $\Omega_\mathrm{M,0}=0.3$, $\Omega_{\Lambda,0}=0.7$, and H$_0=70$ km s$^{-1}$ Mpc$^{-1}$. In this cosmology, the Universe is 1.145 Gyr old at $z=5.12$, with 1\arcsec\, corresponding to 6.3 kpc. We use AB magnitudes \citep{Oke1983} and all $M_\star$ and star formation rate (SFR) estimates assume a \citet{Chabrier2003} initial mass function (IMF).

\section{Data}
\label{sec:data}

This work is based on data from the Canadian NIRISS Unbiased Cluster Survey (CANUCS; GTO program
\#1208; \citealt{Willott2022}), which consists of observations obtained with the Near Infrared Camera (NIRCam; \citealt{Rieke2023}) and the JWST Near InfraRed Imager and Slitless Spectrograph (NIRISS; \citealt{Doyon2023}). The dataset covers five strong lensing clusters and five flanking fields, where lensing magnification is negligible. 
The survey also incorporates Near Infrared Spectrograph (NIRSpec; \citealt{Jakobsen2022}) PRISM follow-up in all fields and medium resolution spectroscopy using the micro-shutter assembly (MSA) in the MACS1149 flanking field, where our galaxy system is located. 

This field was observed with the NIRCam broadband filters $F090W$, $F115W$,
$F150W$, $F277W$, $F444W$, as well as the medium-band filters $F140M$, $F210M$, $F300M$, $F335M$, $F360M$, $F410M$. The Cycle 2 ``JWST in Technicolor” program (GO program \#3362,
PI: A. Muzzin) added observations in the
$F070W$, $F164N$, $F187N$, $F200W$, $F356W$, $F430M$, $F460M$, and $F480M$ filters. We also obtained imaging with the Advanced Camera for Surveys (ACS) ($F435W$, $F606W$, $F814W$) and the Wide Field Camera 3 (WFC3) (namely $F105W$, $F125W$, $F140W$, and $F160W$) onboard HST.
Further information on the CANUCS/Technicolor reduction, calibration, and photometric products can be found in \cite{Sarrouh2025} and references therein.

Our galaxies were also observed during the Cycle 3 CANUCS follow-up (GTO program 4527; 18 May 2025) with the NIRSpec G395M medium-resolution grating in two MSA configurations, each with an exposure time of $\sim3.1$ ks. 
We processed the data with the STScI \texttt{jwst} stage 1 pipeline, with
custom snowball and $1/f$ noise correction. We then ran the \texttt{jwst} stage 2 pipeline up to the photometric calibration
step and used the \texttt{grizli} and \texttt{msaexp} \citep{Brammer2022,Brammer2023} packages for subsequent analysis. During wavelength calibration, we applied a correction for the known intra-shutter offset along the dispersion direction. 

To avoid self-subtraction and to search for potential spectral signal from each shutter (see Fig.~\ref{fig:cutouts} for the MSA shutter configuration), particularly for tLRD, SAT1, and one of the companion clumps (C3; see Sect.~\ref{sec:basic_prop}), we performed a global background subtraction rather than a drizzled-shutter subtraction.
Since tLRD and BBG spectra were taken on the same MSA exposure and are proximate, we first estimated the background spectrum by fitting a spline to the empty shutters' spectra adjacent to the BBG.
We then used the sky background spectrum for both the BBG and the tLRD background subtraction.
The target SAT1 was observed in a different MSA exposure. We therefore separately estimated the background spectrum from the adjacent shutter spectrum that is not on the C3 companion.

We obtained one-dimensional spectra using a
wavelength-dependent optimal extraction that accounts for the increase in the point-spread function (PSF) full width at half maximum with wavelength. Further details on NIRSpec MSA data processing can be found in \citet{Heintz2025} and \citet{Sarrouh2025}. One of the slits was poorly aligned with the centroid of a source (see Fig.~\ref{fig:cutouts}), so the scaling correction for that spectrum, based on photometry, was larger.

\section{The Stingray}

\subsection{Discovery and basic properties}
\label{sec:basic_prop}

This galaxy group was identified in the CANUCS photometric catalog \citep{Sarrouh2025} and was targeted for NIRSpec follow-up. It is located within an overdensity at $z=5.12\pm0.03$, and its main component, which we call CANUCS-MACS1149-tLRD, shows $F300M$ and $F410M$ excesses consistent with prominent H$\beta$ plus [OIII] and H$\alpha$ emission lines. 

\begin{figure*}[htp]
    \centering
    \includegraphics[width=\linewidth]{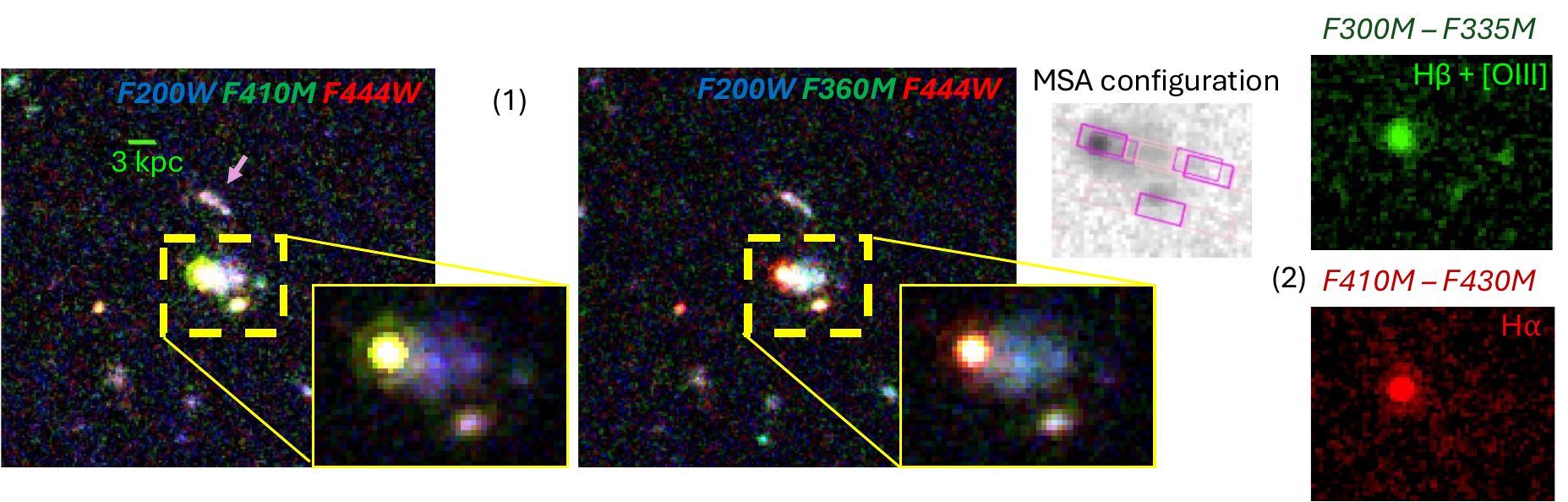}
    \includegraphics[width=\linewidth]{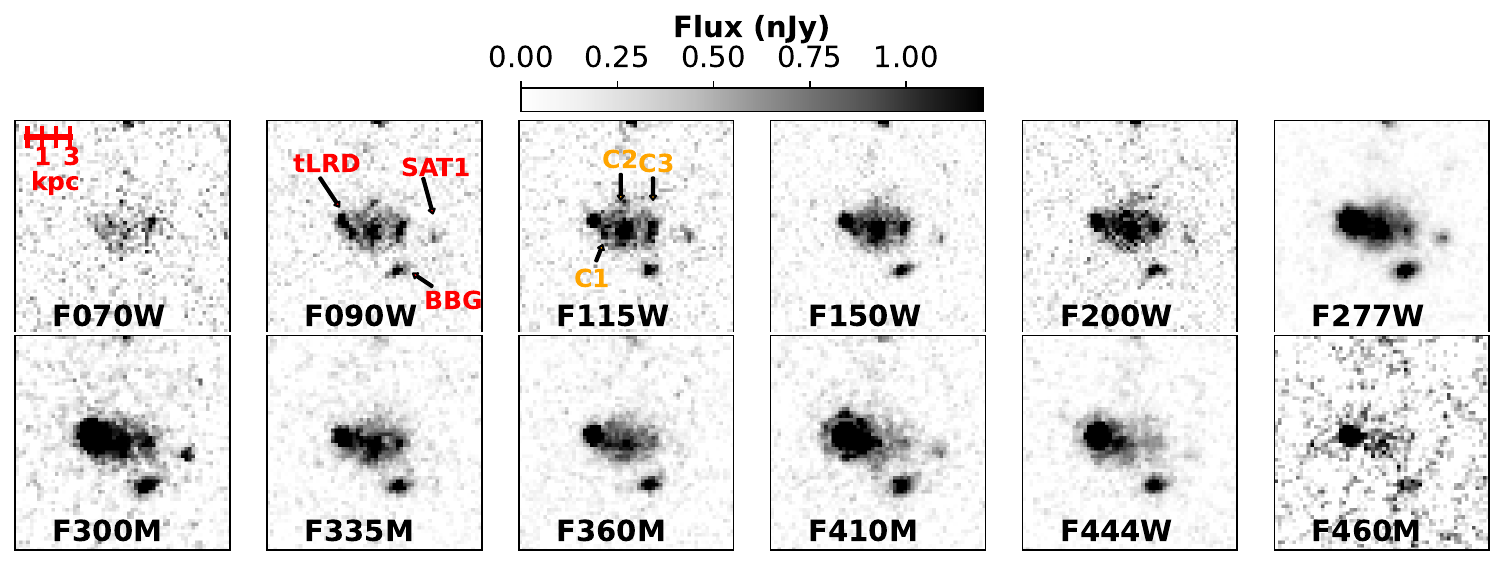}
    \caption{Cutouts of the galaxy group in different bands. Top: (1) 7.7$\times 7.1$ arcsec$^2$ cutouts of two RGB images based on $F200W$, $F410M$, and $F444W$ (left) and $F200W$, $F360M$, and $F444W$ (right), from NIRCam imaging not convolved to the $F444W$ resolution. The yellow square encloses the region that contains the part of the Stingray studied in this work; its dimensions are 2.0$\times 1.4$~arcsec$^2$. A view with a less saturated scale highlights the color differences among the galaxies. These images show the $F410M$ excess in the tLRD (see panel 2), due to H$\alpha$ emission at this redshift, relative to the continuum probed by $F360M$. The pink arrow in the left panel points to a $z\sim1.5$ galaxy group located $\sim1.3$\arcsec\ from the tLRD that resembles the tail of the Stingray. We include a cutout showing the positions of the MSA slits. (2) Cutouts of $F300M - F335M$ (top) and $F410M - F430M$ (bottom) showing the H$\beta$ plus OIII and H$\alpha$ emission, respectively. Bottom: Postage stamps of the sources in different NIRCam bands, not convolved to the $F444W$ resolution. All images are shown at the same scale.}
    \label{fig:cutouts}
\end{figure*}

We nicknamed the group The Stingray. It consists of at least three galaxies at $z=5.12$ for which we have NIRSpec MSA data: CANUCS-MACS1149-tLRD, -BBG, and -SAT1 (see labels in Fig.~\ref{fig:cutouts}). We derived the $z_{\mathrm{spec}}$ values for all three galaxies based on the H$\alpha$ emission lines in their spectra. The tLRD and BBG sources are separated by 5.1~kpc; the BBG and SAT1 are separated by 2.7~kpc; and SAT1 and the tLRD are 3.8~kpc apart.

We used \texttt{SExtractor} \citep{Bertin1996} on the $F150W$ band to recover the centroids of potential sources in the vicinity. This procedure identified three companions, C1, C2, and C3, along with a fuzzy structure centered at C2. Another galaxy group lies to the north (pink arrow in Fig.~\ref{fig:cutouts}), $\sim1.3$\arcsec\, from the tLRD. This northern galaxy group is likely a $z\sim1.5$ system, based on photometric redshifts derived with \texttt{EAzY} \citep{Brammer2008}. Only photometric data are available for these sources.

\begin{table}[]
\caption{Coordinates and redshifts of the galaxies presented in this work.}
    \centering
    \begin{tabular}{c|c|c|c|c}Source&RA &dec &photo-$z$&spec-$z$\\
    \hline
        tLRD & 11:49:33.732 & +22:16:06.892 & 5.21 & 5.1245\\ \hline
        BBG & 11:49:33.688 & +22:16:06.352 & 5.26 & 5.1258\\ \hline
        SAT1 &11:49:33.656  &  +22:16:06.717& 5.21 & 5.1238\\ \hline
        C1 & 11:49:33.725 & +22:16:06.754 & 5.22 & -\\ \hline
        C2 & 11:49:33.708 & +22:16:06.813 & 0.80 & -\\ \hline
        C3 & 11:49:33.686 &+22:16:06.838  & 0.78 & -\\ \hline
    \end{tabular}
    \label{tab:coordinates}
\end{table}

\subsection{Revised photometry and potential LRD nature}
\label{sec:phot}

Given the proximity of the sources and the presence of the fuzzy emission, we computed photometry in small  0.3\arcsec\, apertures around each source, correcting for Galactic extinction, and accounting for flux losses through aperture corrections. Photometry is based on PSF-matched images homogenized to the $F444W$ resolution. 
We used empirical PSFs, obtained by median-stacking non-saturated bright stars \citep{Sarrouh2024,Sarrouh2025}. 

\begin{figure*}[htp]
    \centering
    \includegraphics[width=\linewidth]{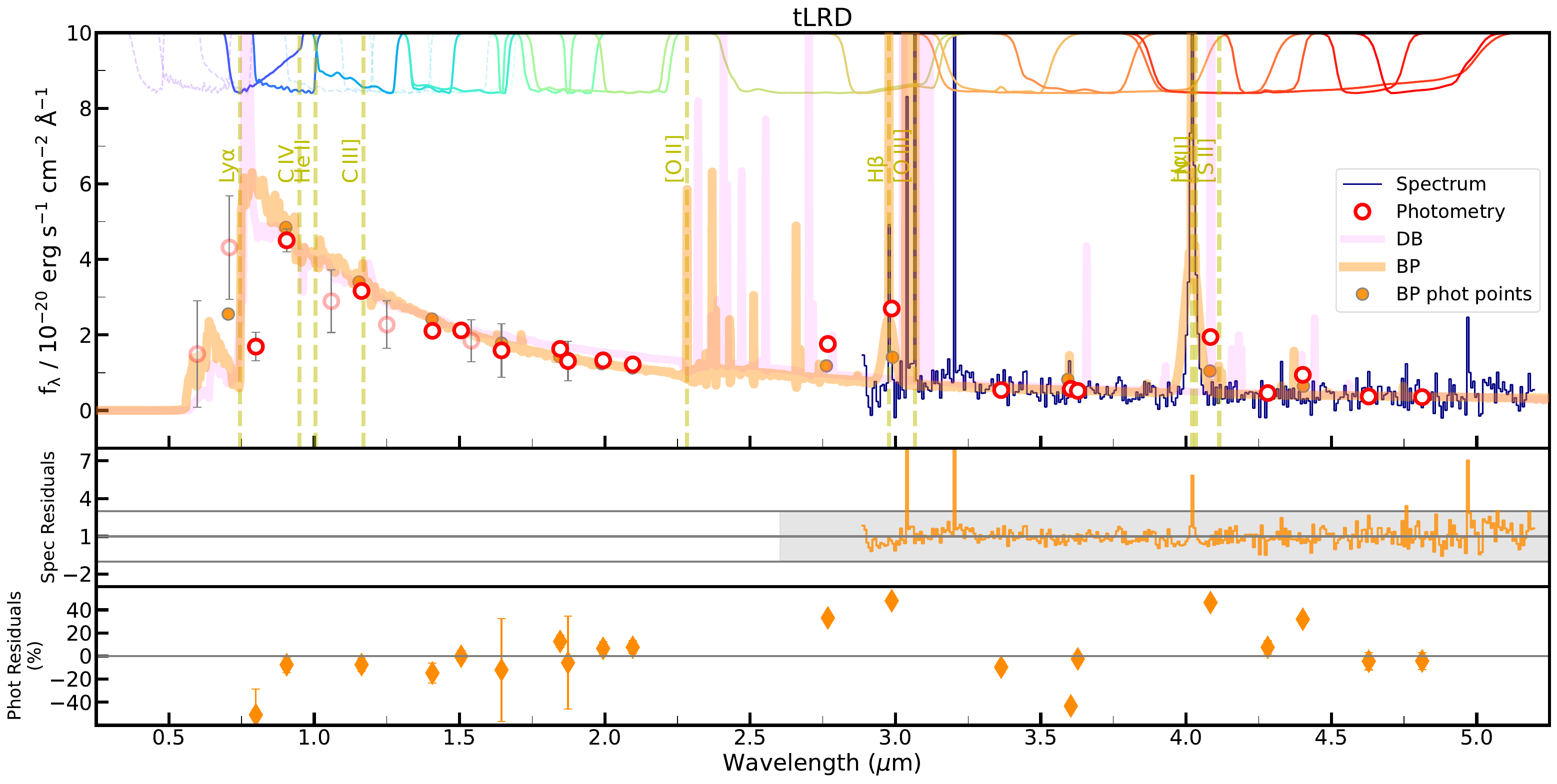}
\caption{G395M MSA spectra (blue) and photometry (open red circles) of the tLRD. The spectrum was rebinned in wavelength to improve S/N and readability. Lighter circles correspond to the HST photometry, which was not considered in our spectro-photometric analysis. The best-fitting models from \texttt{Bagpipes} (spectra plus photometry) and \texttt{DenseBasis} (photometry only) are shown in orange and pink respectively. The \texttt{Dense Basis} fit explored a broader range of redshifts, including the photo-$z$, to ensure convergence, which explains the slight offset between the lines in both models. The photometric points associated with the \texttt{Bagpipes} model are shown as orange circles. We include the ACS and WFC3/HST (dashed) and NIRCam/JWST (solid) filter transmission curves at the top of the panel. Vertical dashed lines highlight the position of the main emission lines at this redshift. We show the \texttt{Bagpipes} residuals underneath. For the spectrum, residuals are computed as the ratio between the spectrum and the model. The shaded gray region corresponds to one standard deviation. For the photometry, we display the difference between the photometry and the model weighted by the photometry. Uncertainties represent the inverse of the S/N. The spectrum dominates the fitting in the optical, resulting in larger residuals in the photometry.}
\label{fig:AGN_spec}
\end{figure*}

\begin{figure*}[htp]
    \includegraphics[width=\linewidth]{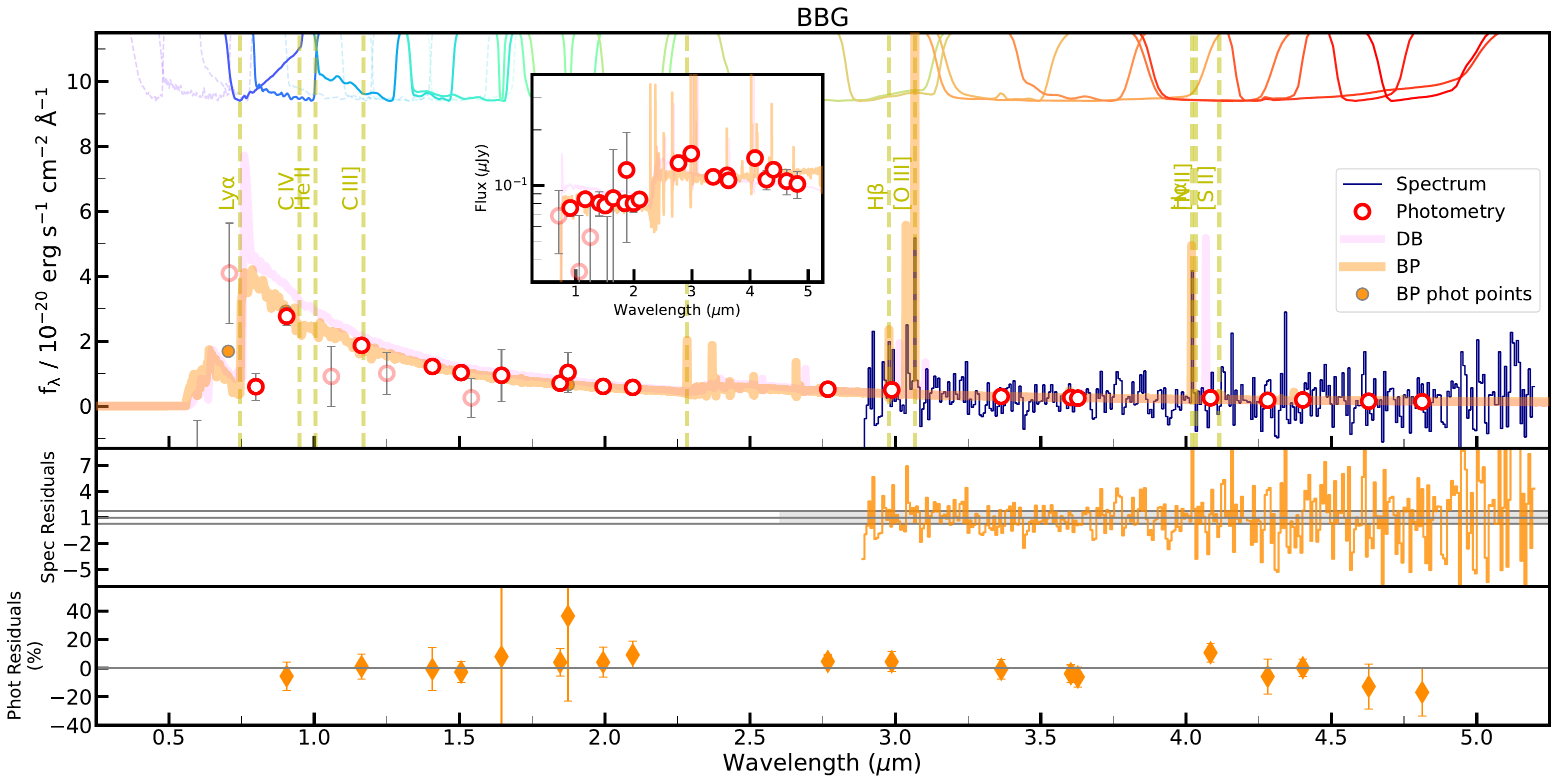}
    \caption{G395M MSA spectra (blue) and photometry (open red circles) of the BBG. We include an inset with the photometry expressed in $\mu$Jy where it is easier to spot the presence of the Balmer break. See Fig.~\ref{fig:AGN_spec} for a complete description of the markers and color codes used in this plot.}
    \label{fig:BBG_spec}
\end{figure*}

\begin{figure*}[htp]
    \includegraphics[width=\linewidth]{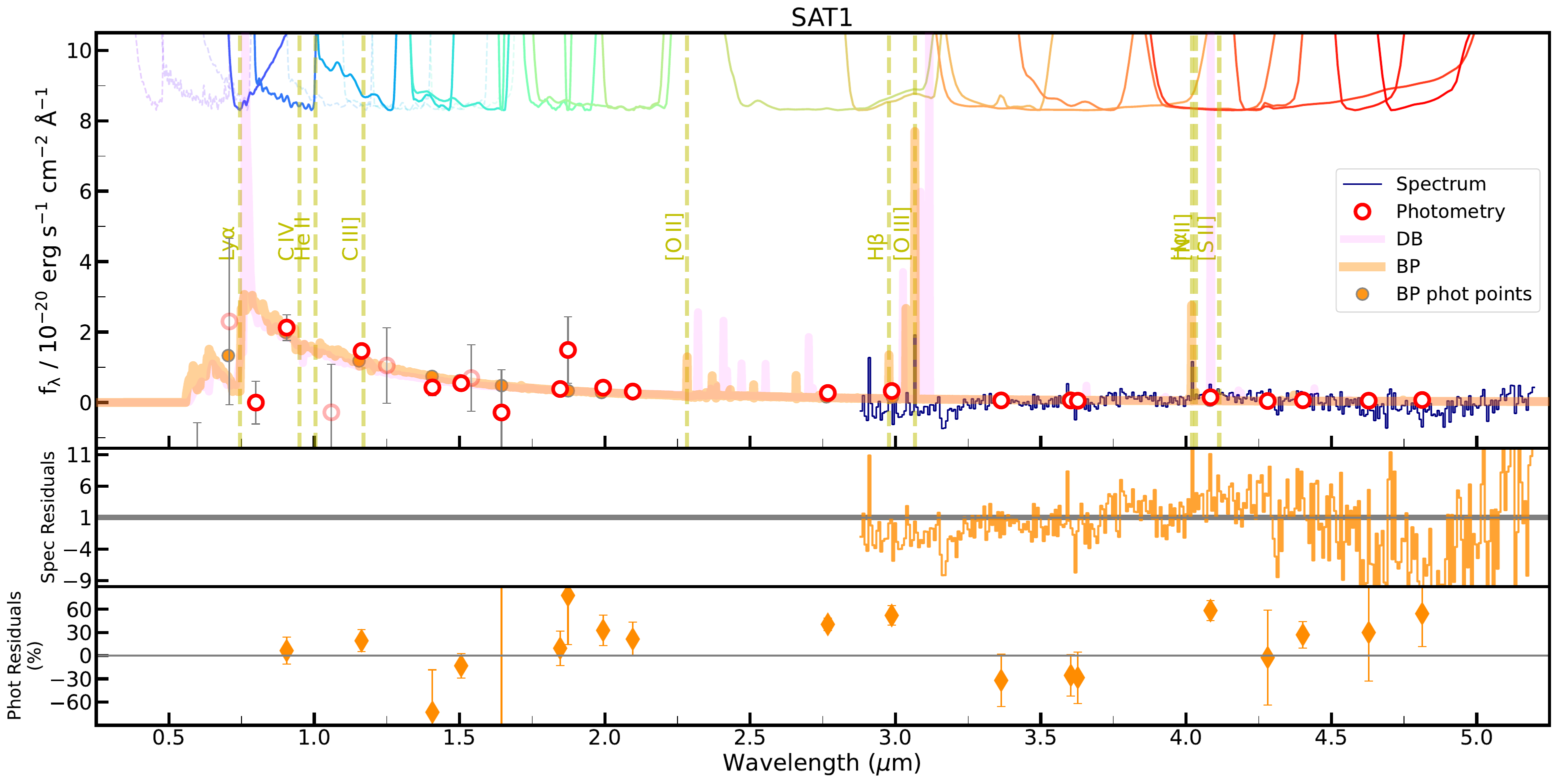}
    \caption{Same as Fig.~\ref{fig:AGN_spec}, but for the SAT1.}
    \label{fig:SAT1_spec}
\end{figure*}

Aperture corrections are based on the ratio between the flux enclosed within 2.5$\times R_{\mathrm{Kron}}$ \citep{Kron1980} and the 0.3\arcsec\, aperture photometry in the $F200W$ band, which is not susceptible to contamination from emission lines at this redshift. We derived uncertainties following the procedure described in \citet{Sarrouh2025},  based on the flux measured in 2,000  0.3\arcsec\, circular apertures placed in blank regions of the sky.
In Appendix~\ref{app:fuzz}, we also consider an additional correction to test potential fuzz contamination in the tLRD photometry and SED-derived properties. Figs. \ref{fig:AGN_spec}, \ref{fig:BBG_spec}, and \ref{fig:SAT1_spec} show the photometric points for the tLRD, BBG, and SAT1, respectively.
We ran \texttt{EAzY} to derive photometric redshifts (photo-$z$s) for all sources, using the standard templates augmented with the \citet{Larson2023} set and the intergalactic medium attenuation curve of \citet{Asada2024photoz}. 

\begin{table}[htp]
\caption{Criteria values from \citet{Kokorev2024} and \citet{Kocevski2024} computed for the tLRD.}
    \centering
    \begin{tabular}{c|c}
    Criterion&Value\\ \hline
    \multicolumn{2}{c}{Kokorev+24}\\
    \hline
    $F115W - F150W < 0.8$ mag & 0.13 mag\\ 
    $F200W - F277W > 0.7$ mag & 1.00 mag\\
    $F200W - F356W > 1.0$ mag & \textbf{0.36 mag}\\
    $bd: F115W - F200W > -0.5$ mag & 0.24 mag\\
    $f_{f444w}(0.5\arcsec)/f_{f444w}(0.3\arcsec) < 1.7$ & 1.33\\\hline
    \multicolumn{2}{c}{Kocevski+24}\\ \hline
    S/N$_{F444W} > 12$ &249\\
    $\beta_\mathrm{opt} > 0$&$-$\textbf{1.36}\\ 
    $-2.8 < \beta_{\mathrm{UV}} < -0.37$&$-1.59$ \\
    $r_{\mathrm{h}} < 1.5 r_{\mathrm{h,\,stars}}$&1.72 pix\\
    $\beta_{F277W - F356W} > -1$&$-$\textbf{4.35} \\ 
    $\beta_{F277W - F410M} > -1$& 0.26\\
    \hline
    \multicolumn{2}{c}{Optical slope from spec is $-$\textbf{0.41}$^*$}\\ \hline
    \end{tabular}
    \tablefoot{The numbers in bold highlight the criteria that tLRD does not satisfy; these are related to its optical emission. The \citet{Kokorev2024} compactness criterion refers to the ratio between the flux measured in 0.4\arcsec and 0.2\arcsec\, apertures. The symbol $bd$ indicates brown dwarf; $r_\mathrm{h}$ is the half-light radius measured in the $F444W$ of \citealt{Kocevski2024}). In this case $1.5 r_{\mathrm{h,\,stars}}\sim2.6$~pix. In the \citet{Kocevski2024} criteria, $\beta = \frac{0.4\,(\mathrm{m_1-m_2})}{\mathrm{log}(\lambda_2/\lambda_1)}-2$. Following Table 2 in that paper, the UV filters correspond to $F115W$, $F150W$, and $F200W$, and the optical filters to $F277W$, $F356W$, and $F444W$. Although $\beta_\mathrm{opt} < 0$ when derived using $F277W$ and $F444w$, $\beta_\mathrm{opt} = 2.1$ when using $F356W$ and $F444W$. We also include the $\beta_\mathrm{opt}$ value estimated directly from the spectrum.\\$^*$ This is the only reliable measurement of the optical slope, given that $F277W$, $F410M$, and $F444W$ are contaminated by strong emission lines.}
    \label{tab:lrd}
\end{table}

The companion C1 can be identified only in the short-wavelength bands in Fig.~\ref{fig:cutouts}. The image resolution and its proximity to the tLRD make its analysis challenging. The companions C2 and C3, along with the fuzzy structure, are likely $z\sim0.8$ projections. In Appendix~\ref{app:companions}, we present the photometry and SED fitting of C1, C2, and C3. In our analysis, we consider only the tLRD, BBG, and SAT1, for which we have spectroscopic confirmation. Table~\ref{tab:coordinates} lists the position of all sources, along with their photometric and spectroscopic redshifts.

The presence of a broad H$\alpha$ line in the tLRD spectrum, its blue rest-frame UV slope, and its compactness may indicate potential LRD activity in this galaxy. We computed the LRD photometric criteria from \citet{Kocevski2024} and \citet{Kokorev2024} and found that the rest-frame UV slope is compatible with LRD emission, but the rest-frame optical slope does not reproduce the ``V''-shape criteria. The optical slope derived directly from the spectrum is larger than that derived from the photometry (which is contaminated by the presence of strong emission lines in $F277W$, $F410M$, and $F444W$), but it is still insufficient to pass the LRD screening. We report the values computed for each criterion in Table~\ref{tab:lrd}. In Sec.~\ref{sec:discussion2}, we discuss a possible connection between this source and LRDs, as this galaxy system may represent a transitional stage between a traditional AGN and an LRD. 

\section{Physical properties}

\subsection{Emission line properties}
\label{sec:lines}

\begin{figure}[htp]
    \centering
    \includegraphics[width=\linewidth]{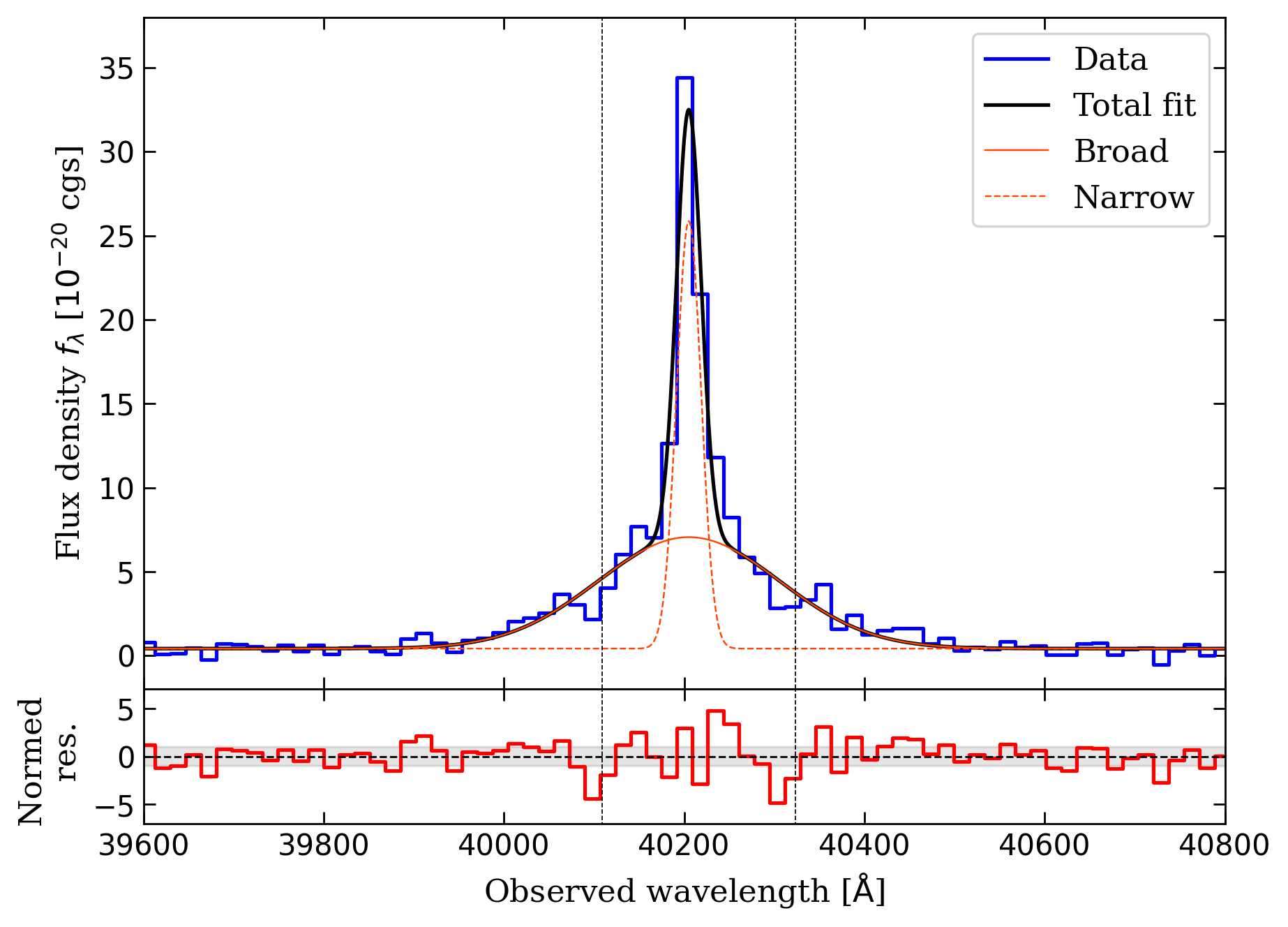}
    \caption{H$\alpha$ fit (black line) for the tLRD compared with the data (blue). The spectrum was rebinned in wavelength to improve the S/N and readability. The solid (dashed) red line shows the fit to the broad (narrow) component. The normalized residuals (spectrum $-$ model, normalized by the error) are displayed underneath. The shaded gray region corresponds to $\pm$1$\sigma$.}
    \label{fig:profile}
\end{figure}

Figs. \ref{fig:AGN_spec}, \ref{fig:BBG_spec}, and \ref{fig:SAT1_spec}. show the G395M MSA 1D spectra of the tLRD, BBG, and SAT1, respectively. We measured the H$\beta$, [OIII]$\lambda\lambda 4959, 5007$, and H$\alpha$ emission using the Levenberg–Marquardt nonlinear least-squares algorithm. For the H$\alpha$ measurement of the tLRD, we selected the trust region reflective algorithm, which provides bound support, for fitting the different line components. The H$\beta$ emission from this object can be fit using a single narrow component.
Figure~\ref{fig:profile} shows the fit to the H$\alpha$ profile of the tLRD source, and Table~\ref{tab:fitting} lists the luminosities derived for each line and object. We find no evidence for a broad component in the H$\alpha$ emission of the BBG and SAT1.

We estimated the AGN bolometric luminosity  from the fluxes of these lines. 
We computed L$_{\mathrm{bol}}$ based on the broad H$\alpha$ flux, assuming the \citet{Greene2005} relation and the bolometric correction from \citet{Richards2006}. \citet{Netzer2019} presented a calibration based on the narrow H$\beta$ component, while \citet{Heckman2004} used the [OIII] line to estimate L$_{\mathrm{bol}}$ from a sample of type 2 AGN.
The [OIII] calibration yields the largest estimate, L$_{\mathrm{bol}} = 1.45\pm0.05 \times 10^{46}$~erg/s. These values decrease to 4.01$\pm0.15 \times 10^{45}$~erg/s and 0.96$\pm0.05 \times 10^{45}$~erg/s when using the H$\beta$ and H$\alpha$ lines, respectively. 

The black-hole mass ($M_{BH}$) is derived from the width of the H$\alpha$ broad component using the empirical relation given by \citet{Reines2013}. We obtain log $M_{BH} = 7.46 \pm 0.09\,M_\odot$, which places our source above the $M_{BH}$ versus $M_\star$ scaling relation, consistent with the $4<z<11$ AGN sample of \citet{Maiolino2024}.

The Eddington ratios are 3.89$\pm0.21$ from [OIII], 1.08$\pm$0.06 from H$\beta$, and 0.26$\pm$0.02 from H$\alpha$. These values imply that, depending on the calibration, the tLRD may be emitting at sub-Eddington to super-Eddington levels. We adopted the values derived from the H$\alpha$ calibration as our standard reference, since we cannot resolve the broad H$\beta$ line. We note that all of these narrow lines are susceptible to contamination from star formation (indeed, [OIII] is resolved in the NIRCam imaging), which would lead to an overestimation of the ratios. Our H$\alpha$-based estimate is compatible with Eddington ratios reported for LRDs in the literature (\citealt{Lin2024} and references therein).

\subsection{Spectrophotometric fitting}
\label{sec:fitting}

We used \texttt{Bagpipes} \citep{Carnall2018} to perform a spectrophotometric fitting of the tLRD, BBG, and SAT1. This \texttt{Python} software utilizes a Bayesian inference approach and allows fitting of the UV and optical emission from an AGN accretion disk using a double power-law model. The code also implements nonparametric SFHs using a series of piecewise constant functions in lookback time. We used the \texttt{continuity} mode, based on the nonparametric SFH formulation of \citet{Leja2019}.

All fits were performed using the Multinest sampling algorithm, and the priors adopted for each parameter are listed in Table~\ref{tab:priors} in Appendix~\ref{app:priors}. The emission from HST filters was not considered in our analysis because of their low signal-to-noise ratio (S/N). We fit the spectrum and photometry of the tLRD with \texttt{Bagpipes} using a composite model consisting of a stellar population plus an AGN. For BBG and SAT1, only stellar populations were included. The redshift was fixed to the spectroscopic values. 

The AGN continuum emission was modeled as a broken power law characterized by three parameters: two spectral slopes and the continuum flux at the break point (5100 $\AA$). The AGN line emission was included as a single H$\alpha$ line. The intensity was left free, while the line width was fixed to 1,773 km/s, as measured from the spectrum. Nebular emission was incorporated through the ionization parameter U, and we assumed a \citet{Calzetti2000} attenuation law.

In addition, we also used the \texttt{Dense Basis} code (\citealt{Iyer2017}, \citealt{Iyer2019}) to fit the photometric data of all galaxies and compare the results with the \texttt{Bagpipes} output. We ran \texttt{Dense Basis} selecting a nonparametric SFH. Stellar population synthesis models are incorporated through the flexible stellar population synthesis (\texttt{FSPS}; \citealt{Conroy2010}) \texttt{Python} module. Contributions from AGNs at UV-to-optical wavelengths cannot be included, so the code therefore assumes that all the emission is of stellar origin. Table~\ref{tab:priors} lists the adopted priors.

We therefore adopted a traditional approach based on a composite of AGN and stellar emission to study tLRD. As discussed in Sects.~\ref{sec:phot} and \ref{sec:discussion2}, this object does not satisfy the optical criteria for LRDs. According to the current BH$^*$ model, the optical continuum emission is associated with a dense gas envelope surrounding the central engine (e.g., \citealt{Inayoshi2024}, \citealt{deGraaf2025}, \citealt{Naidu2025}). If this envelope were present in the tLRD, an additional component would be required in the fitting. However, tLRD likely no longer hosts this envelope, given its flatter optical continuum. This, as further discussed in Sect.~\ref{sec:discussion2}, may indicate an early step in the transition towards a normal AGN or an LRD, characterized by the absence of the gas envelope. An unobscured, broad-line AGN model is therefore the most adequate for the tLRD, given that the general LRD population does not present dust emission in the infrared bands (e.g., \citealt{Casey2024}, \citealt{Setton2025}). 

Furthermore, a UV and optical decomposition of our data would be inaccurate since we cannot spectroscopically probe the V-shaped inflection point (see, however, \citealt{deGraaf_review}, who fit the gas envelopes of a sample of 36 LRDs).

Figures \ref{fig:AGN_spec}, \ref{fig:BBG_spec}, and \ref{fig:SAT1_spec} show the best-fitting models from both codes overlaid on the data and Table~\ref{tab:fitting} lists the values for the different physical parameters. The two codes yield consistent results, indicating that the tLRD and BBG are $\sim10^9\,M_\odot$ sources, while SAT1 is the less massive source in the group ($\sim10^8\,M_\odot$). Although \texttt{Dense Basis} does not consider AGN emission, the $M_\star$ is only slightly overestimated compared with the value reported by \texttt{Bagpipes}. The tLRD host exhibits the highest level of attenuation, approximately $ 0.7$ mag. Overall, the properties of this tLRD are relatively straightforward to reconcile with traditional models, in contrast to many LRDs (e.g., A$_v>1-2$ mag, $M_\star >10^{10}\,M_\odot$; see \citealt{Gentile2024}, \citealt{Tripodi2024}, \citealt{Merida2025}). In terms of their SFRs, the tLRD shows the highest value, followed by the BBG, and lastly SAT1. These values, together with the SFHs (see Sec.~\ref{sec:sfhs}), reflect the star-forming nature of the system, consistent with the presence of emission lines in their spectra.

\begin{figure}[htp]
    \centering
    \includegraphics[width=\linewidth]{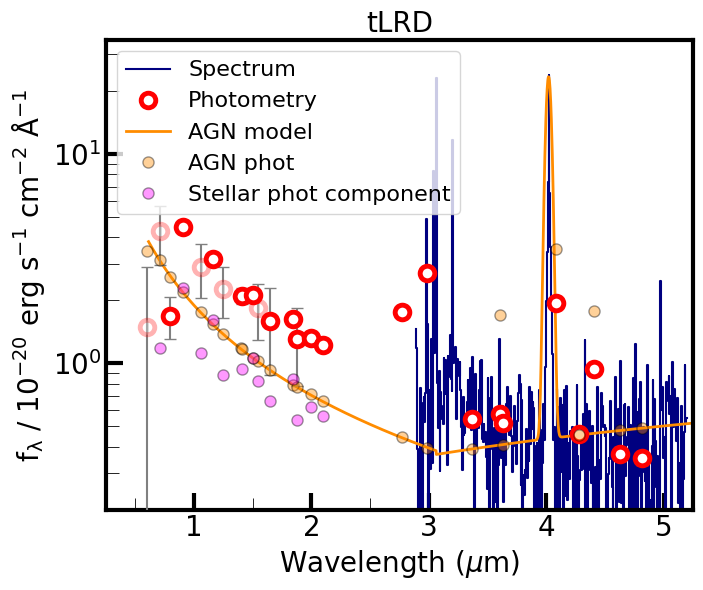}
    \caption{Spectrum (blue) and photometry (open red circles) of the tLRD source, together with the AGN model extracted from \texttt{Bagpipes} (orange). The photometric points associated with the model are shown as orange circles. The photometry linked to the stellar component, obtained by subtracting the AGN model photometry from the observed photometry, is displayed as fuchsia circles. In the optical, the AGN model photometry lies above both the photometry and the photometric points derived from the spectrum.}
    \label{fig:agn_model}
\end{figure}
\begin{figure*}[htp]
    \centering
    \includegraphics[width=\linewidth]{stingray_sfhs.pdf}
    \caption{Star formation histories of the tLRD (left), BBG (middle), and SAT1 (right), obtained using \texttt{Bagpipes} (orange;  based on the \citealt{Leja2019} approach) and \texttt{Dense Basis} (cyan). The shaded regions indicate the 16th-84th percentile range.}
    \label{fig:sfhs}
\end{figure*}

We also derived the AGN model from the best-fitting composite \texttt{Bagpipes} template, shown in Fig.~\ref{fig:agn_model}. The median parameters defining the broken power law are $\alpha_\lambda = -1.43_{-1.74}^{-1.17}$, $\beta_\lambda = 0.64_{0.24}^{1.05}$, $\texttt{hanorm} = 2.24_{2.17}^{2.26}\times10^{-17}$ erg/s/cm$^2$, and $f_{5100} = 3.72_{2.40}^{5.25}\times10^{-21}$ erg/s/cm$^2$/Hz. The parameters $\alpha_\lambda$ and $\beta_\lambda$represent the slopes on either side of 5100$\AA$. The $\alpha_\lambda$ slope for this AGN is consistent with the slopes measured for Sloan Digital Sky Survey (SDSS) quasars \citep{VandenBerk2001}, spanning $0.044\leq z \leq 4.789$. The continuum is normalized at 5100$\AA$ using the $f_{5100}$ parameter, while \texttt{hanorm} defines the total flux of the H$\alpha$ line.

\begin{table*}[htp]
\caption{Best-fitting parameters from \texttt{Bagpipes} (spectrum and photometry) and \texttt{Dense Basis} (photometry only).}
\small
\renewcommand{\arraystretch}{1.5}
\renewcommand{\tabcolsep}{0.1cm}
    \centering
    \begin{tabular}{c|c|c|c||c|c|c||c|c|c|c}
         ID&log $M_\star/M_\odot$& log SFR [$M_\odot$/yr] & A$_v$ [mag] &log $M_\star/M_\odot$& log SFR [$M_\odot$/yr] & A$_v$ [mag] & L$_{\mathrm{H}\alpha}$& L$_{\mathrm{H}\beta}$& L$_{\mathrm{[OIII]\lambda 4959}}$& L$_{\mathrm{[OIII]\lambda 5007}}$\\ \hline
         &\multicolumn{3}{c||}{\texttt{\large{\textbf{Bagpipes}}}} &\multicolumn{3}{c||}{\texttt{\large{\textbf{Dense Basis}}}}&\multicolumn{4}{c}{ [$10^{41}$ erg/s]}\\ \hline
         
         tLRD & 9.08$_{9.03}^{9.14}$ & 1.09$_{1.04}^ {1.15}$& 0.72$_{0.63}^{0.80}$&9.25$_{9.25}^{9.25}$& 1.14$_{1.16}^{1.17}$&0.82$_{0.82}^{0.83}$&82.78$\pm$1.86&7.93$\pm$0.85&13.20$\pm$0.85&41.47$\pm$1.32\\ 
         & && && &&B:54.52$\pm$1.69&&& \\
         & && && &&N:28.26$\pm$1.07&&& \\ \hline
         BBG &9.02$_{8.95}^{9.09}$&0.73$_{0.61}^{0.83}$ &0.28$_{0.14}^{0.40}$&9.10$_{8.94}^{9.22}$&0.52$_{0.45}^{0.89}$&0.03$_{0.02}^{0.09}$&8.14$\pm0.98$&$<3.26$&3.58$\pm$0.81&9.97$\pm1.06$\\ \hline

         SAT1 & 7.84$_{7.66}^{8.02}$&$-0.27_{-0.35}^{0.13}$& 0.07$_{0.03}^{0.13}$&8.28$_{8.06}^{8.60}$ & 0.10$_{-0.02}^{0.22}$&0.11$_{0.03}^{0.23}$&2.19$\pm$0.36&$<1.15$&1.43$\pm$0.31&4.24$\pm$0.40\\ \hline
    \end{tabular}
    
    \label{tab:fitting}
    \tablefoot{We include the 16th and 84th percentiles as upper and lower bounds for the SED-derived properties. The star formation rates correspond to SFR$_{100}$. \texttt{Dense Basis} assumes that all emission originates from stars, whereas \texttt{Bagpipes} includes AGN emission in the fit. In column 8, B and N denote broad and narrow components, respectively. The H$_\alpha$ line width is 1,773 km/s. The BBG and SAT1 H$\beta$ emission line measurements are 2$\sigma$ upper limits.}
\end{table*}

The photometric points associated with this AGN model are also shown, along with the photometry of the stellar component, which was estimated by subtracting the AGN photometry from the observed photometry. This approach assumes that any remaining emission is powered entirely by stars. According to \texttt{Bagpipes}, the rest-frame UV photometry is a composite of stellar and AGN emission, with $\sim56$\% being supplied by the AGN. 
Optical emission would be primarily driven by the AGN, although there is likely a significant contribution from stars powering [OIII] plus H$\beta$ emission (see Appendix~\ref{app:tests}). A red stellar component, outshone by the AGN, may also contribute to the optical (\citealt{Narayanan2024}, \citealt{Whitler2023}). 

\subsection{Star formation histories}
\label{sec:sfhs}

Figure \ref{fig:sfhs} shows the SFHs of the three galaxies, obtained with \texttt{Bagpipes} and \texttt{Dense Basis} using a nonparametric approach (see Sec.~\ref{sec:fitting}). Both codes yield consistent results, indicating that all three galaxies are star-forming at $z=5.12$.

The figure also shows that the SFHs of the  tLRD and SAT1 have similar shapes but differ in scale, with the tLRD exhibiting higher SFRs. Both display rising SFHs, which became steeper $\sim10$~Myr before the epoch of observation.
In contrast, the BBG shows a more complex SFH according to \texttt{Dense Basis}, with continuous star formation until $\sim250$ Myr ago, when its SFR began to rise. Both codes indicate that this galaxy reached an SFR peak, after which the SFH flattened. This flattening could lead to a quenching event, as observed for the $z=5.20$ BBG reported by \citealt{Strait2023}, or the BBG's SFH could revert to a continuous or steeper trend.
In Sec.~\ref{sec:discussion1}, we further analyze these trajectories to investigate the role of interactions in the evolution of these galaxies.

\begin{figure}[htp]
    \centering
\includegraphics[width=\linewidth]{stingray_mass_assembly.pdf}
    \caption{
    Top: Mass assembly histories of the tLRD (solid orange), BBG (solid red), and SAT1 (solid cyan) shown on a logarithmic scale. The dashed lines in the same colors indicate the mass growth of these galaxies, assuming that they follow the main sequence. To compute the MS tracks, we used the main sequence fit from \citet{Merida2025_ms} at $5<z\leq 7$ and the $M_\star$ of the sources at $t = 500$ Myr. We assumed a return fraction of 0.3. Vertical lines highlight $t = 10$ Myr (violet) and $t = 100$ Myr (lime). Bottom: SFHs of our sources, following the same color code. The SFHs of analogous galaxies evolving along the main sequence since $t=500$ Myr ago are also depicted. The dotted lines display the MS evolution resetting the seed mass to that achieved in each time bin.
    The shaded region shows the SFH uncertainty of each source. The scatter of the main sequence, which is $\sim0.2$ dex at this redshift, should also be considered.}
    \label{fig:mass_history}
\end{figure}

\section{Discussion}
\label{sec:discussion}

\subsection{Mass growth induced by the interaction}
\label{sec:discussion1}

In this $z_{\mathrm{spec}}=5.12$ galaxy system, three components exhibit rising SFHs during part of their evolution and show signs of ongoing star formation. The source SAT1 has a lower mass, whereas the other two sources currently have comparable $M_\star$ of $10^{9}\,M_\odot$. One galaxy hosts an AGN, while the other exhibits a Balmer break. 
The three objects lie close together on the sky and share the same redshift. Assuming a system size of 10~kpc (see Fig.~\ref{fig:cutouts}) and assuming a relative velocity of 100 km/s (\citealt{Lambas2003}, \citealt{Ellison2011}), the crossing time for this group is $\sim100$~Myr. This estimate is consistent with the key timescales inferred by the SFHs (see Fig.~\ref{fig:sfhs}), suggesting that interactions between the three galaxies could play a role in their star formation histories. We aimed to determine whether the galaxies' growth results from secular processes or is driven by environmental effects. 

To examine how the galaxies in the Stingray system evolved, we first estimated their mass-assembly histories.
Figure~\ref{fig:mass_history} shows the mass growth of the tLRD, BBG, and SAT1, derived by integrating their \texttt{Bagpipes} SFHs while assuming a return fraction of 0.3 (i.e., as expected for a \citealt{Chabrier2003} IMF). Approximately $500$~Myr ago, the BBG was the most massive source, with $M_\star = 10^{8.2}\,M_\odot$. The tLRD had a mass of $10^{7.4}\,M_\odot$, whereas SAT1 was the least massive object, with $10^{6.7}\,M_\odot$. The BBG increased its $M_\star$ by $\sim0.8$~dex during the last 500 Myr, while the tLRD grew  by $\sim1.5$ dex and SAT1 by $\sim1$~dex. The tLRD and SAT1 assembled a larger fraction of their $M_\star$ than BBG over the same 500 Myr period. This is especially true for the tLRD, which showed a significant steepening of its mass growth $\sim100$~Myr ago. 

To understand whether these differences in the mass growth are due to interactions, as well as to the presence of an AGN in the tLRD, we compared the observed mass growth to that of hypothetical sources that evolved along the star-forming main sequence (MS). By comparing the fiducial MS galaxies with our observations, we can infer whether the growth observed in our three galaxies can be attributed to secular processes or if a contribution from other factors is required.
To calculate the reference MS histories, we adopted the initial $M_\star$ of our three objects at $t=500$~Myr as the seed masses. We then evolved them over time, using the parameters of the MS relation from \citet{Merida2025_ms} for galaxies at $5<z\leq7$. 

The results of this comparison are summarized below. The BBG mass assembly history is consistent with the evolution along the MS, whereas the final mass of the tLRD is $\sim0.7$ dex higher than  expected for an MS galaxy. The tLRD started deviating from the MS trend 100~Myr ago. The growth of SAT1 is more difficult to interpret. This galaxy was a very low-$M_\star$ source that may have been in a latent or lulling phase 500~Myr ago (see \citealt{Looser2024}, \citealt{Mintz2025}). However, a potential latent state in SAT1 cannot be confirmed because MS studies at this redshift are incomplete at such low $M_\star$, preventing us from modeling a reference MS galaxy for SAT1 with confidence. We note, however, that the final mass of SAT1 is roughly consistent with that of an MS galaxy, considering its $M_\star$ at $t = 500$~Myr.

Figure~\ref{fig:mass_history} also shows the SFHs of the sources, alongside the SFHs of the reference MS model galaxies. In this figure, dotted curves show hypothetical SFHs for three objects that follow the same mass assembly histories as our galaxies. In these models, the seed mass is reset to the $M_\star$ achieved in each time bin rather than being set once at 500~Myr. 
The BBG displays several upturns in its SFH, but all are consistent with MS evolution, particularly when accounting for the typical MS scatter of $\sim0.2$~dex at this redshift. The tLRD SFH is consistent with MS tracking up to $\sim100$~Myr ago, when it experienced a star formation upturn that displaced it by $\sim1$~dex above the MS. About 10~Myr ago, the tLRD experienced a new upturn that elevated it by 1.5~dex above the MS, or by 1~dex if the MS is reset at $t = 100$~Myr. The combination of these two bursts allowed the tLRD to equalize the BBG in $M_\star$. The SFH of SAT1 falls slightly below the MS at early times, although it remains consistent with it within the SFH uncertainties. Its subsequent evolution is also consistent with the MS within uncertainties until 10 Myr ago, when the galaxy experienced an upturn that moved it $\sim1.5$ dex above the MS.

In summary, the evolution of BBG can be explained by secular evolution along the MS. During this period, the galaxy developed a Balmer break, while stars continued to form at a rate consistent with that expected for an MS galaxy. The BBG is not bursting, as also reflected by the strength of its emission lines relative to the UV continuum. In contrast to the BBG, the tLRD experienced two bursts of star formation that cannot be explained by evolution along the MS. These events occurred 10 Myr and 100 Myr ago. Lastly, the evolution of SAT1 is also consistent with the MS, except for a recent burst of star formation that began 10~Myr ago. Although starbursts can have origins other than interactions, the proximity of these sources makes the environment a likely precursor of these events. 

Given the configuration of all three galaxies at the time of observation and the estimated crossing times, we propose a scenario in which the tLRD encountered the BBG 100 Myr ago. This interaction mainly boosted the star formation in the tLRD, which was the less massive source by 0.6 dex -- or a factor of four -- in mass at that time. The BBG may also have experienced some increase in star formation as a result of the encounter, but this was not sufficient to push the galaxy into a starburst phase. About 10 Myr ago, SAT1 joined the system and interacted with the BBG, which lies closer in projected distance. Again, this effect was more noticeable in SAT1, given that its $M_\star$ is 1.5 dex lower than the masses of the other two galaxies, corresponding to a factor of  $\sim$30. The 10~Myr upturn in the tLRD could be attributed to an interaction with SAT1, although 10~Myr ago the $M_\star$ of the tLRD was already comparable to that of BBG, and it was located further away from SAT1. 

Alternatively, this 10 Myr burst in tLRD could be due to feedback from its AGN. Activation of the AGN can be delayed with respect to the triggering of star formation (see \citealt{Emonts2006}, \citealt{Omori2025}, \citealt{ORyan2025}). In realistic hydrodynamic models, AGN luminosities do not
peak at the same time as the gas density on small scales, but rather at later times, when most of the gas has been consumed by star formation and/or expelled by stellar feedback \citep{Hopkins2012}. Gravitational torques, which govern further inflow into the center, are relatively inefficient in gas-dominated systems. However, stars provide an efficient angular momentum sink, allowing for a more rapid inflow \citep{Hopkins2012}. This process can result in a delay of $\sim100$~Myr between the onset of star formation and subsequent AGN activation. 
Thus, delayed AGN triggering may be a consequence of the 100 Myr burst induced by the tLRD$-$ BGG encounter. Feedback from AGN would then be the mechanism responsible for triggering a new star formation upturn in tLRD $\sim10$ Myr ago.

We acknowledge that this discussion is based on the assumption that the SED-derived properties of these galaxies, particularly the SFHs, are reliable. This assumption is particularly important for the tLRD. Although the tLRD can be well fit with \texttt{Bagpipes}, it exhibits LRD features that pose a challenge for current models. To assess the robustness of our results, we tested alternative solutions by varying the AGN component priors. We then examined the impact on the SFHs. These results are presented in Appendix~\ref{app:tests}, where we show that trends persist despite changes in the AGN priors. 

Finally, it is worth highlighting the broader connection of the Stingray system to the emerging framework of enhanced hierarchical galaxy formation at high redshift. This group resembles the building blocks system (BBS) discussed in Asada et al. (in prep), comprising a BBG and two low-mass star-forming galaxies at $z=5.2$, and is likely destined to evolve into a Milky Way-mass galaxy by the present day. The BBS is undergoing ``boosted hierarchical assembly,'' in which the $M_\star$ of the resulting merger product is not merely the sum of the preexisting masses of the individual components, but is also dramatically boosted by interaction-induced bursts of star formation. In the case of the Stingray's tLRD, $M_\star$ appears to have increased fourfold through the interaction, prior to its anticipated future merger with the BBG.

Although more massive than the BBS, our galaxy system provides another key ingredient to the boosted hierarchical assembly scenario: AGN activity. This activity can potentially enhance the effects of the environment, extending the interaction-induced star formation burst through delayed AGN activation. This process can lead to even more rapid mass assembly in these galaxies. Moreover, our observations suggest that the environment can act as a precursor of black hole mass growth in low-$M_\star$, high-$z$ objects. 

\subsection{Transitional LRDs}
\label{sec:discussion2}

\begin{figure}[h!]
\centering
\includegraphics[width=.93\linewidth]{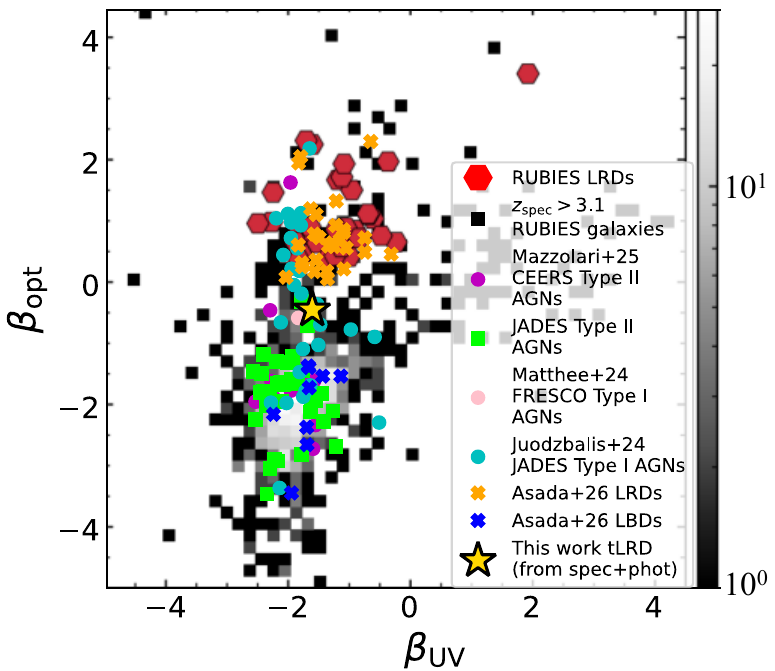}
\includegraphics[width=.95\linewidth]{asada_uv_halpha.pdf}
    \caption{Adapted versions of Fig. 6 (top) from \citet{Hviding2025} and Fig. 3 (bottom) from \citet{Asada2026} showing $\beta_{\mathrm{opt}}$ vs. $\beta_{\mathrm{UV}}$ and UV luminosity $\nu \mathrm{L}\nu$ vs. the broad $\mathrm{L}_{\mathrm{H}\alpha}$ for a sample of LRDs and non-LRDs. Top: We compared tLRD (yellow star) with the LRDs selected in RUBIES (red hexagons) and \citet{Asada2026} (orange crosses).
    \citet{Hviding2025} included $z_{\mathrm{spec}} > 3.1$ galaxies from RUBIES as 2D grayscale histograms. We added sources from \citet{Mazzolari2025} (magenta circles); Type II AGNs from JADES (lime squares) \citep{Scholtz2025}; a Type I AGN from \citet{Matthee2023} (pink circle); and Type I AGNs from \citet{Juodzbalis2025} (cyan circles), all extracted from \citet{Hainline2024arxiv}. Little Blue Dots (LBDs) from \citet{Asada2026} are represented as blue crosses. Bottom: tLRD (yellow star) is compared to the LRDs and LBDs (red and blue points) from \citet{Asada2026}. We included the typical relation for local Type I AGNs (solid line; \citealt{VandenBerk2001}) and young starburst galaxies (dashed line; from BPASS+Cloudy photoionization modeling). The $\mathrm{L}_{\mathrm{H}\alpha}$ for tLRD was derived from the spectrum (see Table \ref{tab:fitting}), selecting the broad component. $\nu \mathrm{L}\nu$ was estimated from the photometry. Both quantities were corrected for dust attenuation following \citet{Asada2026}.}
    \label{fig:lrd_hints}
\end{figure}

The AGN in the Stingray shows properties that are compatible with an LRD phase. It is compact and blue in the rest-frame UV (see Sec.~\ref{sec:phot}) and shows a broad H$\alpha$ component (Sec.~\ref{sec:lines}). However, the optical emission does not satisfy the LRD screening criteria. The AGN is likely to be emitting in the sub-Eddington regime (Sec.~\ref{sec:lines}), in line with LRDs reported in previous studies.

The top panel of Fig.~\ref{fig:lrd_hints} is an adaptation from Fig. 6 of \citet{Hviding2025}, showing the position of our source in the $\beta_{\mathrm{opt}} - \beta_{\mathrm{UV}}$ plane. We compare the properties of tLRD with those of the LRDs of \citet{Hviding2025}, extracted from the Red Unknowns: Bright Infrared Extragalactic Survey (RUBIES; \citealt{deGraaf2025Rubies})\, and the LRDs from \citet{Asada2026}. These objects display a spectroscopic ``V''-shaped continuum, compact sizes, and broad Balmer lines. The combination of the spectrum and the photometry places our AGN at the intersection between the LRD population and the $z_{\mathrm{spec}} > 3.1$ RUBIES galaxies. 

\citet{Hainline2024arxiv} also explored the $\beta_{\mathrm{opt}} - \beta_{\mathrm{UV}}$ plane (see Fig. 4 therein). Near the UV and optical slope values of the tLRD, we find a fraction of Type 1 AGNs from \citet{Juodzbalis2025}, one Type 1 AGN from \citet{Matthee2023}, one Type II AGN from \citet{Mazzolari2025}, and one Type II AGN from the \textit{James Webb} Advanced Deep Extragalactic Survey (JADES; \citealt{Eisenstein2023}). There is a significant fraction of high-$z$ AGNs that exhibit much bluer $\beta_{\mathrm{opt}}$. However, most Type I AGNs are located within the LRD locus. This could suggest that Type I AGNs are mostly compatible with LRDs or with sources similar to the tLRD. Nevertheless, \citet{Hainline2024arxiv} note that LRDs are easy to select (as they are more luminous) and therefore easy to follow up spectroscopically. So far, Type I AGNs at $z > 6$ are mostly consistent with LRDs, likely as a result of the difficulty in selecting normal Type I AGNs at these redshifts. This may imply that the tLRD could instead be a normal luminous blue Type I AGN, rather than the result of a transition phase.

In the second panel of Fig.~\ref{fig:lrd_hints}, we show the position of our source in the UV luminosity $\nu \mathrm{L}\nu$ vs. the broad $\mathrm{L}_{\mathrm{H}\alpha}$ plane. This figure is adapted from Fig. 3 in \citet{Asada2026}. In that paper, they selected LRDs with PRISM spectra from the DAWN JWST Archive \citep{brammer2025} by looking for broad H$\alpha$ line emitters at $5 < z_{\mathrm{spec}} < 7.2$ with rest UV continuum detection, and imposing the LRD criteria following \cite{Hviding2025}. They then selected a sample of Little Blue Dots (LBDs) by relaxing the criterion imposed on the optical slope (i.e., $\beta_\mathrm{opt} - \beta_\mathrm{UV} < 0.5$). These galaxies would thus be compact and blue, resembling LRDs, but their optical slope would not pass the LRD screening (note that $\nu \mathrm{L}\nu$ values are not available for the Type I AGNs from \citealt{Hainline2024arxiv}). These LBDs are also included in the top panel of Fig.~\ref{fig:lrd_hints}. The position occupied by tLRD in the bottom panel of Fig.~\ref{fig:lrd_hints} is consistent with the region in between the LRD and LBD regimes, as defined by the typical relation for Type I AGN and young starburst galaxies.

Considering these results together with the properties reported previously, it is possible that this object was caught in a transition phase between normal AGNs and LRDs. In other words, at some point an LRD evolves with redshift to a state that no longer meets the selection criteria, implying a transition phase during which these galaxies retain LRD features while simultaneously developing new characteristics \citep{Billand2025}. The question then arises whether this tLRD is evolving into a regular LRD or whether its LRD phase is fading.

\citet{Billand2025} investigated the possible evolution of LRDs by identifying candidates in Cosmic Evolution Early Release Science (CEERS; \citealt{Finkelstein2017}, \citealt{Finkelstein2025}) data that could represent descendants of LRDs. They report that the decline in the number density of LRDs over cosmic time is driven by the acquisition of a stellar component that settles in the outskirts of the galaxy. As $M_\star$ increases, the characteristic ``V''-shape fades and the
physical size of the galaxies grows. Similarly, \citet{Hainline2024arxiv} propose a trend between luminosity and optical slope, whereby lower-luminosity AGNs display bluer slopes as a consequence of the star-forming host galaxy becoming more dominant
in these weak AGNs.

\citet{Billand2025} propose several scenarios, including mergers, for the transition from LRDs to traditional AGNs. However, their results indicate that cold accretion is the leading mechanism.
In their study, galaxies were selected to retain a red optical core. These galaxies exhibit rest-frame UV slopes that do not meet the ``V''-shape criteria and have sizes inconsistent with being compact, unlike the tLRD. 

A different scenario for this transition is proposed by \citet{Kido2025} within the BH$^*$ framework. They suggest that a decrease in the infalling mass onto the BH envelope can cause a decrease in its mass (i.e., when the infall rate is lower than the Eddington accretion rate). Combined with BH mass growth, this process leads to the dissolution of the gas envelope. According to \citet{Inayoshic}, supernova explosions from massive stars in nuclear starbursts can inject sufficient energy and momentum to expel gas from the nucleus, quenching the gas supply to the BH envelope and ultimately driving a transition to a normal AGN phase. The LRD stage would only persist for one-third of a Salpeter time. This scenario could explain the features observed in the tLRD, which no longer exhibits a red optical slope. However, such short timescales for the transition imply that this phase may be almost undetectable. Thus, the possibility that the tLRD is a blue Type I AGN should be considered.

The existence of channels for LRD disruption also implies the presence of inverse channels. For example, a gas envelope could form through super-Eddington accretion; for this to occur, the mass infall rate from the interstellar medium must significantly exceed the BH’s Eddington rate. Otherwise, radiative and mechanical feedback from the BH would suppress the infall \citep{Kido2025}. 

The LRD state may be a phase that can be activated and deactivated (although above a certain critical $M_{BH}$, subsequent LRD phases are unlikely; \citealt{Inayoshic}), probably linked to the AGN's accretion rate, which is itself directly related to the burstiness of the system. The environment can greatly influence the activation and deactivation of the LRD phase (see also \citealt{Merida2025}). Within a certain cosmic time window, environmental conditions facilitate the emergence of LRDs, while at lower redshifts, this channel becomes increasingly compromised.

It remains unclear whether the tLRD will move towards a more traditional AGN or instead evolve into a regular LRD, as the nature of these objects is not yet fully understood.
Our results suggest that this transition phase (regardless of its direction) could be linked to an environmental process that began $\sim100$~Myr ago, although the conditions in this system are very particular. To fully understand the LRD phase and its underlying mechanisms, it is essential to study a larger and more complete sample of these transitional-like objects that exhibit mixed features of LRDs and normal AGNs.

\section{Summary and conclusions}

In this work, we present The Stingray, a galaxy system at ${z_{\mathrm{spec}} = 5.12}$ identified in an overdensity within the MACS 1149 flanking field from the CANUCS observations. We analyzed the properties of three galaxies in this system, dubbed the tLRD, the BBG, and SAT1, which appear as an interacting system. Similarly to the building blocks system presented by Asada et al. (in prep.), this system provides an example of ``boosted hierarchical galaxy assembly.''

The SFH of the tLRD and SAT1 show evidence of interaction processes that took place 10 and 100 Myr ago, linked to the BBG. These interactions led to a more rapid growth of stellar mass in the tLRD and SAT1 than that expected from secular processes, which are dominant in the BBG. Additionally, the 10 Myr burst in tLRD can be attributed to a delay in AGN activation relative to the star formation burst. A later activation could extend the interaction-driven burst that took place in the tLRD 100~Myr ago. Consequently, the environment allows for black hole mass growth in such low-$M_\star$ high-$z$ systems.

The AGN in the Stingray exhibits properties that are partially compatible with LRDs. The tLRD may therefore represent a transition object (see \citealt{Billand2025}), capturing the emergence or the fading out of an LRD phase at $z_{\mathrm{spec}} = 5.12$, with galaxy-galaxy interactions acting as a potential trigger.

\begin{acknowledgements}
This research was enabled by grant 18JWST-GTO1 from the 
Canadian Space Agency and Discovery Grant and Discovery Accelerator funding 
from the Natural Sciences and Engineering Research Council (NSERC) of Canada to MS.\\
This research used the Canadian Advanced Network For Astronomy Research (CANFAR) platform operated in partnership by the Canadian Astronomy Data Centre and The Digital Research Alliance of Canada with support from the National Research Council of Canada, the
Canadian Space Agency, CANARIE, and the Canada Foundation for Innovation.\\
This work is based on observations made with the NASA/ESA/CSA James Webb Space Telescope. The data were obtained from the Mikulski Archive for Space Telescopes at the Space Telescope Science Institute, which is operated by the Association of Universities for Research in Astronomy, Inc., under NASA contract NAS 5-03127 for JWST. JWST observations are associated with programs JWST-GTO-1208, -4527, and GO-3362 (doi:10.17909/ph4n-6n76, doi:10.17909/cyh7-mm53,
doi:10.17909/18nv-np70).
\end{acknowledgements}

%
%

\bibliography{aa57594-25}
\bibliographystyle{aa}

\begin{appendix}

\section{The fuzz}
\label{app:fuzz}

Centered around C2, and covering C1, C3, and partially the tLRD, we can see a fuzzy structure that could potentially affect the measurements of all these galaxies. In this appendix, we discuss the properties of this fuzzy emission and include a correction intended to remove potential contamination in the tLRD photometry that might not be accounted for by the aperture correction.

To define this fuzzy structure, we measured its emission in residual images computed for each band. These residual images were obtained using the PSF photometry functions from \texttt{photutils}. We modeled the emission of the sources in the system with \texttt{psfphot}, which provided us with the residuals of the PSF fit. The aperture radius, used to estimate the initial flux of each source, was set to 3 pixels in \texttt{psfphot}, with \texttt{fit$\_$shape}~=~(5, 5). We used the PSF-matched images homogenized to the $F444W$ resolution.

PSF photometry assumes that the sources are point-like in each of these bands, which is not the case here. This implies that our residual images may also include emission from the aforementioned objects. The latter could lead to an overestimation of the fuzzy emission. However, this is not concerning since the purpose of this Appendix is to provide some insights into the fuzz features, as well as a range within which the SED-inferred properties of the tLRD can vary.

We used the \texttt{voronoi$\_$2d$\_
$binning} method from the \texttt{vorbin} \texttt{Python} package in the residual images. This package is an implementation of the two-dimensional adaptive spatial binning method of \citet{Cappellari2003}. It utilizes Voronoi tessellations to bin data to a given minimum S/N, allowing us to define the regions contributing to the fuzz. We computed pixel-by-pixel photometry of each of these regions and combined them to produce the SED of the total fuzz. The photo-$z$ was estimated using \texttt{EAzY}, and the SED-derived properties were computed as in Sec.~\ref{sec:fitting}. We show the photometry, best-fitting models, and SFH of the fuzz in Fig.~\ref{fig:fuzz_fit}. According to our analysis, this is a $z\sim0.8$ projection of $10^8\,M_\odot$, with A$_V$ = 0.07 mag, and log SFR [$M_\odot$/yr] $= -1.45$. It could be dynamically coupled with C2 and/or C3, at the same redshift.

\begin{figure}[htp]
    \centering
\includegraphics[width=\linewidth]{stingray_appendix_fuzz_sed.pdf}
\includegraphics[width=\linewidth]{stingray_appendix_fuzz_sfh.pdf}
    \caption{Top: photometry of the fuzzy emission (open red circles) and best-fitting models from \texttt{Bagpipes} and \texttt{Dense Basis} (in orange and violet). See Fig.~\ref{fig:AGN_spec} for a complete description of the markers and color codes used in this plot. Bottom: SFHs derived with \texttt{Bagpipes} (orange) and \texttt{Dense Basis} (cyan). The shaded regions correspond to the area enclosed within the 16th and 84th percentiles.}
    \label{fig:fuzz_fit}
\end{figure}

To compute the correction for the tLRD photometry, we drew three one-pixel-width annuli, separated by half a pixel, around the 0.3\arcsec\, aperture. Looking at Fig.~\ref{fig:cutouts}, we see that the tLRD is off-center from the fuzz. Part of the tLRD is not exposed to this emission, so we divided the annuli in half, assuming the left side is free from contamination. This allows us to estimate the fuzz flux with respect to the sky in each annulus. We took the mean value considering the three annuli. This was done for every filter, resulting in a more significant correction for the rest-frame UV than for the optical. However, this correction is small enough that the tLRD retains its blue UV slope. We show the effects of this correction on the tLRD photometry in Fig.~\ref{fig:fuzzcorr}. The SED-derived parameters from \texttt{Bagpipes} based on the corrected photometry are log $M_\star/M_\odot = 8.98_{8.85}^{9.09}$, A$_V = 0.86_{0.77}^{0.94}$ mag, and log SFR$[M_\odot/$yr] = $0.95_{0.86}^{1.06}$. These are compatible with the estimates reported in Table~\ref{tab:fitting}, and the SFHs also match. Therefore, the presence of this fuzzy emission does not alter our results.    

\begin{figure}[htp]
    \centering
\includegraphics[width=\linewidth]{stingray_appendix_fuzz_corr.pdf}
    \caption{Effect of the fuzz correction on the tLRD photometry. The original photometry is shown as open red circles while the corrected values are displayed as open blue circles.}
    \label{fig:fuzzcorr}
\end{figure}

\section{The companions: C1, C2, and C3}
\label{app:companions}

We used \texttt{Bagpipes} and \texttt{Dense Basis} to fit the SEDs of C1, C2, and C3. Their photometry, best-fitting models, and SFHs are displayed in Fig.~\ref{fig:companions}. Table~\ref{tab:fitting_app} comprises their physical properties. C1 is the only source that could be at $z\sim5$ according to these codes. We observe a dropout from $F090W$ to $F070W$, which is consistent with the Lyman break at $z\sim5$. There is also a hint of [OIII] emission at $z\sim5$ in the $F300M-F335M$ image (see Fig.~\ref{fig:cutouts}). However, the contamination due to the tLRD makes its fitting and interpretation quite challenging.

\begin{figure*}[htp]
    \centering
    \includegraphics[width=\linewidth]{stingray_appendix_sed.pdf}
   \includegraphics[width=\linewidth]{stingray_appendix_sfh.pdf}
    \caption{Top: Photometry and SED fitting of C1 (left), C2 (middle), and C3 (right). See Fig.~\ref{fig:AGN_spec} for a complete description of the markers and color codes used in this plot. Bottom: SFHs of these objects, following the same order.}
    \label{fig:companions}
\end{figure*}

\begin{table*}[]
\caption{Best-fitting parameters of the companions derived using \texttt{Bagpipes} and \texttt{Dense Basis}.}
\renewcommand{\arraystretch}{1.5}
    \centering
    \begin{tabular}{c|c|c|c|c|c|c}
         ID&log $M_\star/M_\odot$& log SFR [$M_\odot$/yr] & A$_v$ [mag] &log $M_\star/M_\odot$& log SFR [$M_\odot$/yr] & A$_v$ [mag]\\ \hline
         &\multicolumn{3}{c|}{\texttt{Bagpipes}} &\multicolumn{3}{c}{\texttt{Dense Basis}}\\ \hline
         
         C1 & 8.69$_{8.65}^{8.75}$ & 0.70$_{0.67}^ {0.74}$& 0.61$_{0.57}^{0.65}$&9.04$_{9.04}^{9.24}$& 0.70$_{0.70}^{0.98}$&0.43$_{0.43}^{0.49}$\\ \hline

         C2 &8.00$_{7.90}^{8.13}$&$-0.75_{0.91}^{-0.60}$ &0.14$_{0.05}^{0.25}$&8.10$_{7.53}^{8.38}$&$-0.71_{-0.97}^{-0.56}$&0.23$_{0.07}^{0.46}$\\ \hline

         C3 & 8.45$_{8.36}^{8.57}$&$-0.93_{-1.19}^{-0.72}$& 0.23$_{0.08}^{0.42}$&8.50$_{8.32}^{8.61}$ & $-1.13_{-2.22}^{-0.74}$&0.22$_{0.06}^{0.46}$\\ \hline
    \end{tabular}
    
    \label{tab:fitting_app}
    \tablefoot{We include the 16th and 84th percentiles. The star formation rates correspond to SFR$_{100}$.}
\end{table*}

\section{\texttt{Bagpipes} and \texttt{Dense Basis} priors}
\label{app:priors}

In Table~\ref{tab:priors} we show the values for the priors used in \texttt{Bagpipes} and \texttt{Dense Basis}. 

\begin{table}[htp]
\caption{Priors used in this work.}
    \centering
    \begin{tabular}{c|c}
    \multicolumn{2}{c}{\texttt{Bagpipes}}\\ \hline
     SFR:    &  \\
    log $Z/Z_\odot$ & (0.001, 2.5)\\
    log $M_{\star, \mathrm{formed}}/M_\odot$& (6,13)\\
    Bin edges [Myr] & (0, 10, 100, 250  \\
    & 500, 800, 1100)$^*$  \\ \hline
    A$_V$ [mag] & (0, 2)  \\ \hline
    log U & (0, $-4$)  \\ \hline
    AGN: & \\
    $\alpha_\lambda$ & ($-2.5$, 3) \\
    $\beta_\lambda$ & ($-1.5$, 2.3) \\ \hline
    \multicolumn{2}{c}{\texttt{Dense Basis}}\\ \hline
    log $M_{\star}/M_\odot$& (6, 13)\\ \hline
    Dust prior & Flat \\ \hline
    sSFR prior & sSFRflat\\ \hline
    log sSFR & ($-12,-7.5$)\\
    \hline
    
    \end{tabular}
    
    \label{tab:priors}
\tablefoot{$^*$The bin edges shown in the table were used for the $z=5$ sources, namely the tLRD, BBG, SAT1, and C1. For C2, C3, and the fuzz, we included bins at 2500, 5000, and 5500 Myr. We set a \citet{Calzetti2000} dust attenuation law in both codes. The redshifts were set to the $z_{\mathrm{spec}}$ (or photo-$z$s in the absence of spectra).}
\end{table}

\section{Testing AGN priors}
\label{app:tests}

A degeneracy between the AGN model parameters, $A_V$, and $M_\star$ was already reported in \citet{Tripodi2024} for CANUCS-LRDz8.6.
In order to test the consistency of our SED-derived properties in the case of tLRD, we forced different AGN solutions by changing the $\alpha_\lambda$ and $\beta_\lambda$ priors. The original AGN solution, using broad priors for both parameters (see Table~\ref{tab:priors}), yielded $\alpha_\lambda = -1.43_{-1.74}^{-.17},\,\beta_\lambda=0.64_{0.24}^{1.05}$. The seed and final $M_\star$ are $10^{7.4}\,M_\odot$ and $10^{9.1}\,M_\odot$, respectively.

We selected combinations designed to span distinct regions of the parameter space, including cases where both slopes are negative, both are positive, and one is negative while the other is positive.

As a result of our tests, we got four alternative fits, which we called Tests 1, 2, 3, and 4:

\begin{itemize}
    \item Test 1: $\alpha_\lambda=-2.34_{-2.63}^{-2.07},\,\beta_\lambda=-1.36_{-1.76}^{-0.13}$, $M_{\star,\,\mathrm{seed}} = 10^{7.7}\,M_\odot$, $M_{\star,\,\mathrm{final}} = 10^{9.1}\,M_\odot$
    \item Test 2: $\alpha_\lambda = 0.85_{0.38}^{1.25},\,\beta_\lambda=-1.30_{-1.68}^{-0.83}$, $M_{\star,\,\mathrm{seed}} = 10^{7.9}\,M_\odot$, $M_{\star,\,\mathrm{final}} = 10^{9.1}\,M_\odot$
    \item Test 3: $\alpha_\lambda = 0.78_{0.37}^{1.22},\,\beta_\lambda=1.93_{1.63}^{2.21}$, $M_{\star,\,\mathrm{seed}} = 10^{7.3}\,M_\odot$, $M_{\star,\,\mathrm{final}} = 10^{9.1}\,M_\odot$
    \item Test 4: $\alpha_\lambda = -2.30_{-2.55}^{-2.00},\,\beta_\lambda=1.89_{1.53}^{2.23}$, $M_{\star,\,\mathrm{seed}} = 10^{7.2}\,M_\odot$, $M_{\star,\,\mathrm{final}} = 10^{8.9}\,M_\odot$
    
\end{itemize}

Test 4 was forced to provide larger absolute values of the slopes compared to our fiducial values, obtained using broad priors. Figure~\ref{fig:agn_models_appendix} shows the AGN models for the different tests.

The final $M_\star$ estimates are all consistent with the results provided in Table~\ref{tab:fitting}. The seed masses are slightly less massive in Tests 3 and 4, and more massive in Tests 1 and 2.
We show the corresponding SFHs in Fig.~\ref{fig:sfh_app}.

All tests match the overall trend depicted by the tLRD's SFH, displayed in Fig.~\ref{fig:mass_history}. 
They show larger uncertainties than the original fit for $t>10$~Myr (especially Tests 3 and 4), but reflect upturns at 10 and 100~Myr which cannot be explained by secular evolution alone. Tests 3 and 4 place tLRD below the MS $\sim500$~Myr ago. The upturn taking place 100~Myr ago would have moved tLRD above the MS. 
It is important to remember that these are forced solutions in which we restricted the priors to explore different regimes of the AGN parameters. 

These tests show that, depending on the priors, the relative intensities of the upturns seen in the SFH may vary, but the scenario depicted by the trends remains as described in Sect.~\ref{sec:discussion1}. 100 and 10~Myr ago, tLRD experienced an increase in the star formation likely induced by the interaction and AGN triggering. The final mass of tLRD is $10^9\,M_\odot$. The seed mass is more uncertain, but it is $<10^8\,M_\odot$. 

The fact that the SFH and the $M_\star$ do not vary much when the AGN model assumptions are changed highlights the important role played by the stellar component in the fitting of this galaxy, as also reflected by the strong [OIII] plus H$\beta$ emission in this source (see Figs.~\ref{fig:cutouts} and ~\ref{fig:AGN_spec}).

\begin{figure}
    \centering
    \includegraphics[width=\linewidth]{agn_models_appendix.pdf}
    \caption{AGN models extracted from \texttt{Bagpipes} of the different tests (dashed lines), together with the fiducial model (solid black line) for tLRD.}
    \label{fig:agn_models_appendix}
\end{figure}

\begin{figure*}
    \centering
    \includegraphics[width=0.9\linewidth]{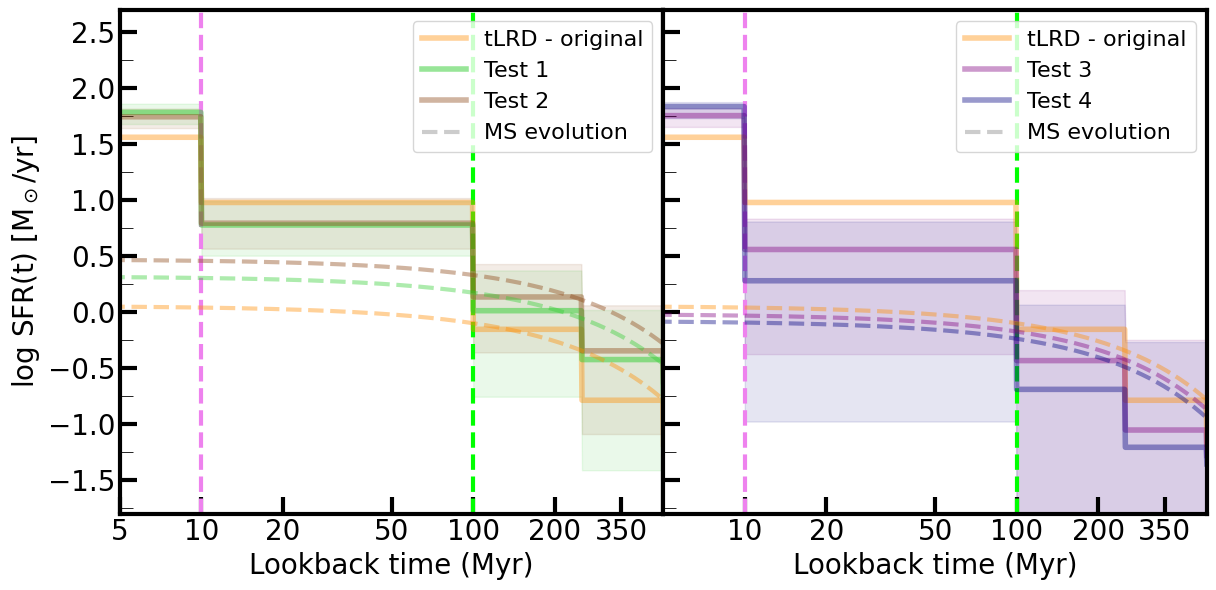}
    \caption{Original SFH of tLRD (solid orange), and SFHs obtained from our tests (green, brown, purple, and navy, following Fig.~\ref{fig:agn_models_appendix} color code) on log scale. We divide the figure in two panels, each showing the results of two tests. SFHs of analogous galaxies evolving along the main sequence since $t=500$ Myr ago are also depicted. The shaded region shows the SFH uncertainty. We do no include the uncertainty for the original SFH solution for the sake of clarity. Vertical lines highlight $t = 10$ Myr (violet) and $t = 100$ Myr (lime), respectively.}
    \label{fig:sfh_app}
\end{figure*}
\end{appendix}

\end{document}